\documentclass[conference]{IEEEtran}

\usepackage{amsmath}
\usepackage{amssymb}
\usepackage{cite}
\usepackage{graphicx}
\usepackage{epstopdf}
\usepackage{url}
\usepackage{cite}
\usepackage{algorithmicx}
\usepackage{algorithm}
\usepackage{{texdef2015}}

\newtheorem{thm}{Theorem}

\newtheorem{definition}{Definition}
\newtheorem{lemma}{Lemma}

\newtheorem{pb}{Problem}

\allowdisplaybreaks

\usepackage{tikz}
\usetikzlibrary{automata,arrows,positioning,calc,fit,shapes.multipart,chains,shapes}
\usepackage{microtype}

\newcommand{\QueueR}[2]{\draw[thick] (#1) ++(-0.7,0.3) --  ++(0.7,0) -- ++(0,-0.6) --  ++(-0.7,0);
 \foreach \i in {1,...,3}
	\draw[thick] (#1) ++(-\i*0.2,0.3) -- ++(0,-0.6);
\coordinate (#2) at ($(#1) + (-0.7,0)$);}

\begin{document}

\title{Cache Updating Strategy Minimizing the Age of Information with Time-Varying Files' Popularities} 
\author{
  \IEEEauthorblockN{Haoyue~Tang, Philippe Ciblat, Jintao Wang, Mich\`ele Wigger,~Roy D. Yates}
}

\maketitle

\begin{abstract}
We consider updating strategies for a local cache which downloads time-sensitive files from a remote server through a bandwidth-constrained link. The files are requested randomly from the cache by local users according to a popularity distribution which  varies over time according to a Markov chain structure. We measure the freshness of the requested  time-sensitive files through their Age of Information (AoI). The goal is then to minimize the average AoI of all requested files by appropriately designing the local cache's downloading strategy. To achieve this goal, the original problem is relaxed and cast into a Constrained Markov Decision Problem (CMDP), which we  solve using a Lagrangian approach and Linear Programming. Inspired by this solution for the relaxed problem, we propose a practical cache updating strategy that meets all the  constraints of the original problem. Under certain assumptions, the practical updating strategy is shown to be optimal for the original problem in the asymptotic regime of a large number of files.
 For a finite number of files, we show the gain of our practical updating strategy over the traditional square-root-law strategy (which is optimal for fixed non time-varying file popularities) through numerical simulations.
 \end{abstract}

\section{Introduction and Problem Statement}
\let\thefootnote\relax\footnotetext{\noindent -----------------\\
  Haoyue Tang and Jintao Wang are with Dept. of Electronic Engineering, Tsinghua University, Beijing, China (\{thy17@mails, wangjintao@mail\}.tsinghua.edu.cn). Philippe Ciblat and Mich\`ele Wigger are with Telecom Paris, Institut Polytechnique de Paris, Palaiseau, France (\{ciblat, wigger\}@telecom-paris.fr). Roy D. Yates is with Rutgers University, New Brunswick, NJ, USA (ryates@rutgers.edu).\\
  This work was supported by the ERC (Grant No. 715111), the National Key R\&D Program of China (Grant No.2017YFE0112300) and Beijing National Research Center
  for Information Science and Technology under Grant BNR2019RC01014
  and BNR2019TD01001.}

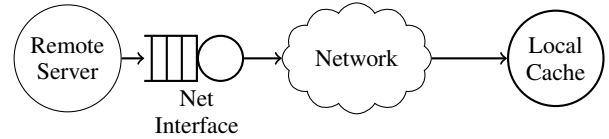
\begin{figure}[htb]
\centering\small
\begin{tikzpicture}[node distance=1cm]
\node [draw,circle, rounded corners,align=center] (newsource) {Remote\\Server};
\node[draw,circle,thick] (NetInterface)[right = 1 of newsource]{\strut};
\node (NetQ) at (NetInterface.west) {};
\QueueR{NetQ}{NetQenter}
\draw node [below = 0.2 of NetQ,align=center] {Net\\ Interface};
\draw[->,thick] (newsource.east) --  (NetQenter.west);
\node  [cloud, draw,cloud puffs=12,cloud puff arc=120, aspect=1.5, inner ysep=1em] [right = 0.5 of NetInterface](net){Network};
\draw[->,thick] (NetInterface.east) --  (net.west);
\node[draw,circle,rounded corners,align=center,minimum height=12mm,thick] (localcache)[right = 1 of net] {Local\\Cache};
\draw[->,thick] (net.east) --   (localcache.west);
\end{tikzpicture}
\caption{Local cache system.}
\label{fig:system}
\end{figure}

Consider a local cache connected by a capacity-constrained  link to a remote network server, as shown in Figure~\ref{fig:system}. The server stores $N$ time-sensitive files that  change in a continuous manner. The local cache maintains a copy of each  file, and, upon request, sends the copy to a local user.  By the capacity-constrained link from the local cache to the server, the cache cannot maintain the latest version of each  item, and the copy it sends to the user can be outdated. The goal of our study  is to measure the freshness of the copies sent to the users in terms of their \emph{age of information (AoI)}, and to propose a cache updating strategy that minimizes the average AoI of  the downloaded copies. Our main focus is on a  setup where the popularities of the various files vary over time. 

Existing works for cache updating with time-sensitive files and time-varying environments mainly aim at minimizing the average AoI when channel states change randomly  \cite{sun17TIT,tang2019minimizing},   energy arrivals are random \cite{arafa_20}, or packet arrivals are random \cite{hsu_scheduling_2018}.  A few works consider time-varying popularities for cache updating, e.g.,  \cite{gunduz_18_tcom,gao,kam_vary}. Similar to our work, \cite{gunduz_18_tcom,gao} propose   policies for downloading files to a local cache over a  capacity-limited link, but their figure of merit is not the AoI. Only  \cite{kam_vary} considers both time-varying popularities and  AoI.

Nevertheless the setting  in  \cite{kam_vary} differs from ours in that in \cite{kam_vary}: i) the cache (as opposed to the  server-cache link) is capacity-limited; ii) the goal is to minimize the missed-files probability in the cache (as opposed to the average AoI), iii)  the current popularity value depends on the past requests weighted by their AoI (as opposed to being determined by a stationary Markov chain independent of the files' AoI as assumed in this paper); and finally iv) the considered AoI is different from ours since their age is defined as the request rate within a given time-interval.
	
	 Our paper can actually be regarded as an extension of our previous work   \cite{yates17isit}  to time-varying file popularities.  It is also related to  \cite{tang20wiopt}, which extended  \cite{yates17isit} to variable update durations.



We now explain our model in more detail. Consider the system in Fig.~\ref{fig:system}, which comprises a remote server holding $N$  files $n=1,\ldots, N$ that are subject to version updates, a local cache downloading the latest versions through a bandwidth limited network, and users requesting files from the local cache. 
 In each time slot $t\in \{1, \ldots, T\}$, the local cache can download the current version of no more than $M$ files from the remote server due to the bandwidth constraint. For each file $n\in\{1,\ldots, N\}$, let  $\{u_{n,t}\}\in\{0,1\}$ be the download decision at time $t$, i.e., $u_{n,t}=1$, if file $n$ is downloaded at time $t$, and $u_{n,t}=0$ otherwise. The bandwidth constraint requires that
\begin{equation}\label{eq:C1}
\sum_{n=1}^Nu_{n, t}\leq M, \quad \forall t. 
\end{equation}

We denote by $X_{n,t}$  the AoI of file~$n$ in slot $t$, i.e.,  the number of slots that have passed since the local cache has downloaded file $n$. For convenience of exposition, we assume that all the files in the local cache are updated at time one so that $X_{n, 1}=1, \forall n$. Afterwards for $t>1$, the AoI evolves as
\begin{align}\label{eq:age-u}
X_{n,t+1}&=
\begin{cases}
1, & u_{n,t}=1;\\
X_{n, t} +1, & u_{n,t} =0.
\end{cases}
\end{align}

In each slot $t$, the number of requests for file $n$ depends on its current popularity mode $R_{n,t}\in \mathcal{R}:=\{1, \cdots, R\}$, where $R>0$. 
 The expected number of requests of file $n$ is determined by a function $\omega_n:\mathcal{R}\rightarrow\mathbb{R}^+$, so that the expected number of requests for file $n$ at time slot $t$ is given by $\omega_n(R_{n,t})>0$. It is assumed that for each file~$n$, the  sequence $\{R_{n,t}\}_{t=1}^T$ evolves according to an $R$-state Markov chain   with transition probabilities $P_{r, r'}^n:=\text{Pr}\left(R_{n, t+1}=r'|R_{n,t}=r\right)$ for $(r, r')\in\mathcal{R}^2$. 


Let $\Pi$ be a set of (cache updating) strategies, such that the design of the downloading decisions $\{u_{n,t}\}$ at slot $t$ only depend on the current and past popularity modes $\{R_{n,\tau}\}_{\tau\leq t}$ and AoIs $\{X_{n,\tau}\}_{\tau\leq t}$ as well as on the  statistics $\{P_{r,r'}^n\}$. The future popularity modes $\{R_{n,\tau}\}_{\tau> t}$ cannot be used. The goal in this article is to design a strategy ${\boldsymbol \pi}\in\Pi$ that minimizes the expected total AoI of all  requested files averaged over an  infinite-time horizon. The corresponding optimization problem can be written as:
\begin{pb}[Original problem]\label{pb:1}
	\begin{subequations}
	\begin{align}
		{\boldsymbol \pi}^*=\arg\min_{{\boldsymbol \pi}\in\Pi}\lim_{T\rightarrow\infty}&\mathbb{E}_{\boldsymbol\pi}\left[\frac{1}{T}\sum_{t=1}^T\sum_{n=1}^{N}\omega_n(R_{n,t})X_{n, t}\right],\\
		\text{s.t. }&\sum_{n=1}^Nu_{n, t}\leq M,\quad \forall t. \label{eq:bandwidth1}
	\end{align}
\end{subequations}
\end{pb}
Problem 1 can be cast into a Markov Decision Process (MDP) where the state contains both the current AoIs $\mathbf{X}_t=[X_{1, t}, \cdots, X_{N, t}]^T$ and popularities $\mathbf{R}_t=[R_{1, t}, \cdots, R_{N, t}]^T$ of all the files, and Eq.~\eqref{eq:bandwidth1} can be cast as a constraint on  the action space $\mathcal{A}:=\{\mathbf{u}_t|\sum_{n=1}^Nu_{n,t}\leq M\}$. The   cardinality of the action space $\mathcal{A}$ of this MDP grows exponentially in $M$, and thus even for moderate    values of $M$ its solution cannot be found using standard algorithms such as  \emph{Relative Value Iteration} \cite{puterman}.

We therefore slightly relax the hard bandwidth-constraint \eqref{eq:bandwidth1} and derive an optimal solution for the relaxed optimization problem, see solution ${\boldsymbol\pi}_R^*$ to Problem 2 in Section~\ref{sec:res}. Motivated by this solution, in Section \ref{sec:algo}  we  propose a practical updating strategy $\hat{\boldsymbol\pi}$ which satisfies the original constraint \eqref{eq:bandwidth1}.  In Section \ref{sec:asymp}, we prove that, under mild conditions, the strategy $\hat{\boldsymbol\pi}$  is optimal for Problem \ref{pb:1} when $N$ goes to infinity for a fixed $N/M$. Numerical illustrations are provided in Section \ref{sec:simus}. Concluding remarks are drawn in Section \ref{sec:ccl}.


\section{Problem Resolution}\label{sec:res}

\subsection{Relaxed problem description}

Similarly to \cite{whittle,yates17isit,hsu_scheduling_2018,igor_18_ton,tang2019minimizing}, we relax Constraint~\eqref{eq:bandwidth1} into an expected infinite-time horizon constraint. This leads to the following relaxed optimization problem: 
\begin{pb}[Relaxed problem]
 	\begin{subequations}
 		\begin{align}
 		{\boldsymbol\pi}_R^*=\arg\min_{{\boldsymbol\pi}\in\Pi}&\lim_{T\rightarrow\infty}\mathbb{E}_{\boldsymbol\pi}\left[\frac{1}{T}\sum_{t=1}^T\sum_{n=1}^{N}\omega_n(R_{n,t})X_{n, t}\right],\label{eq:costfunction}\\
 		\text{s.t. }&\lim_{T\rightarrow\infty}\mathbb{E}_{\boldsymbol\pi}\left[\frac{1}{T}\sum_{t=1}^T\sum_{n=1}^Nu_{n, t}\right]\leq M. \label{eq:bandwidthrelaxed}
 		\end{align}
 	\end{subequations}
 \end{pb}

Problem 2 can be cast into the framework of Constrained Markov Decision Processes (CMDP) \cite{altman1999constrained}. Its  action space however is even larger than that of Problem 1. What makes Problem 2 tractable is that it can be decoupled into independent sub-problems with smaller action spaces, as we now explain.

 The CMDP associated with Problem 2 is a countable-state CMDP with finite set of actions. Consequently, \cite{sennott93} asserts that Problem~2 can be solved, and that the optimal policy ${\boldsymbol\pi}_R^*$ can be determined by introducing the 
 Lagrangian function
\begin{align}
	\mathcal{L}&({\boldsymbol\pi}, W)=\label{eq:Lagrange}\\
	&\lim_{T\rightarrow\infty}\mathbb{E}_{\boldsymbol\pi}\left[\frac{1}{T}\sum_{t=1}^T\sum_{n=1}^N\left(\omega_n(R_{n, t})X_{n, t}+Wu_{n, t}\right)-WM\right]. \nonumber
\end{align} 
associated with the cost function \eqref{eq:costfunction} and the constraint \eqref{eq:bandwidthrelaxed}.

In the remainder of this subsection and the next-following Subsection~\ref{subsec:problem3}, we minimize the Lagrangian for a fixed value of $W$. In Subsection~\ref{subsec:problem2} we then determine the appropriate value(s) of $W$ which lead to the solution of Problem~2.

To minimize the Lagrangian for a fixed value $W$, we first notice that Eq.~\eqref{eq:Lagrange} is separable over the various files. This is easier to see after swapping the order of the summations over $t$ and $n$ in Eq.~\eqref{eq:Lagrange} (the order can be swapped because the sum over $n$ is finite). As a consequence (see \cite[Chapter 4]{stefanov2001} for more details),  
 the minimizing policy ${\boldsymbol\pi}^*_F(W)$ in \eqref{eq:Lagrange} for given $W$ factorizes as
\begin{equation}\label{eq:factorizedpol0}
  {\boldsymbol\pi}^*_F(W)= \bigotimes_{n=1}^N \pi^{n,*}_F(W),
  \end{equation}
  where $\pi^{n,*}_F(W)$ denotes the solution to the following optimization problem for file $n$ only: 
\begin{pb}[per-File relaxed problem]
	\begin{eqnarray}\label{eq:pb3}
	\lefteqn{	\pi_F^{n,*}(W)\triangleq}  \nonumber\\ &&\arg\min_{\pi\in\Pi}\lim_{T\rightarrow\infty}\mathbb{E}_\pi\left[\frac{1}{T}\sum_{t=1}^T\big(\omega_n(R_{n,t})X_{n,t}+Wu_{n,t}\big)\right].\IEEEeqnarraynumspace
	\end{eqnarray}
\end{pb}

\subsection{An algorithmic solution for Problem 3}\label{subsec:problem3}
We solve the optimization problem~\eqref{eq:pb3} for a fixed file index~$n$ and Lagrange multiplier $W$. For simplicity,  we  omit the subscript $n$. $W$ is also omitted except when it appears in a mapping expression. 
 
 Problem 3 can be cast into an MDP with a two-dimensional state $(X_t, R_t)$, action $u_t\in\{0, 1\}$, and instantaneous cost
 \begin{equation}C(X_t, R_t, u_t)=\omega(R_t)X_t+Wu_t. \label{eq:onestepcost}
 \end{equation}
  According to Eq.~\eqref{eq:age-u}, if $u_t=1$, the file is downloaded in slot~$t$ and the AoI drops to $1$ in the next slot; otherwise, the AoI grows by $1$. The state transition relationship thus is
\begin{subequations}
	\begin{align}
		\text{Pr}((X_t, R_t)\rightarrow(1, r))&=P_{R_t, r}, \text{ if } u_t=1;\\
		\text{Pr}((X_t, R_t)\rightarrow(X_t+1, r))&=P_{R_t, r}, \text{ if } u_t=0. 
	\end{align}
\end{subequations}

\begin{definition}
  A  policy $\pi$ is called \emph{stationary}, if for each time $t$ and    AoI-popularity-mode pair $(X_t=x,R_t=r)$,  the action $u_t=1$ is chosen with a probability $\xi_{x,r}$ that only depends on the AoI-popularity-mode pair $(x,r)$ but not on the time $t$.
\end{definition}

The next theorem states  that the optimum solution $\pi_F^*$ is a stationary policy with a specific threshold structure.

\begin{thm}\label{thm:piF}
  There exists an optimal stationary policy $\pi_F^*$  and  a set of thresholds $\{\tau_r\}_{r\in\mathcal{R}}$ such  that  $\pi_F^*$  downloads the file with probability $1$ in state $r$ if $x > \tau_r$ and it keeps idle with probability $1$  if $x<\tau_r$. 
\end{thm}
The proof is similar to \cite[Lemma 1]{tang2019minimizing} but where instead of the  AoI and the channel quality, the  AoI and the popularity mode should be considered for the two-dimensional state. Many other papers prove the optimality of threshold policies, but usually with a one-dimensional state \cite{hsu_scheduling_2018,igor_18_ton}.

By Theorem~\ref{thm:piF}, there exists  a stationary optimal policy such that the AoI is bounded as $X_{\max}= \max_{r\in\mathcal{R}} \tau_r$, where the maximum exists because $\mathcal{R}$ is finite.  Inspecting, \eqref{eq:pb3}, one sees in particular that there must  exist a stationary optimal (threshold) policy with 
\begin{equation}\label{eq:boundXmax}
X_{\max} \leq \max_{r\in \mathcal{R}} \frac{W+ \omega(r)}{\omega(r)}=:X_{\mathrm{ub}}.
\end{equation}
The term $X_{\mathrm{ub}}$ is finite because $\omega(r)>0$ and because $\mathcal{R}$ is finite.
Above inequality \eqref{eq:boundXmax} is obtained by showing that for any policy violating \eqref{eq:boundXmax},  it is possible to find an improved policy that in popularity state $R=r$ updates file $n$ whenever its age  $X> \frac{W+ \omega(r)}{\omega(r)}$. Notice that, in the following subsections, we will often write $X_{\mathrm{ub}}^n(W)$ instead of $X_{\mathrm{ub}}$ to make the dependence on $n$ and $W$ explicit.

We can thus restrict to finding the optimal stationary policy  that solves the CMDP with a restricted \emph{finite} state space $\{1,\ldots, X_{\mathrm{ub}}\} \times \mathcal{R}$. 
Different algorithms for finding  such a policy  are described in \cite{puterman}.  
In this paper, we resort to the \emph{Linear Programming (LP)} approach developed in  \cite{altman1999constrained}, 
and  rewrite Problem 3  in terms of the  two steady-state distributions $\{\mu_{x,r}\}$ and $\{\nu_{x,r}\}$. Here, $\mu_{x,r}$ stands for the steady-state probability of having AoI $X=x$ and  popularity state $R=r$, whereas  the so called \emph{occupation measure} $\nu_{x, r}$ stands for  the steady-state probability of simultaneously having $(X=x,R=r)$ \emph{and} taking the ``download action'' $u=1$.  As a consequence, the probability of updating a file for a given AoI $X=x$ and popularity state $R=r$ equals $\xi_{x,r} = \nu_{x,r}/\mu_{x,r}$. We  set $0/0=1$ by convention. But notice that this  convention has no effect on the solution of the optimization problem because state $(x,r)$ is reached with probability $\mu_{x,r}=0$. With these definitions, and because the cost  $\omega(r)x$ only depends on the state and the constraint $u$ is nonzero only for a single action, we can apply \cite[Theorem 4.3]{altman1999constrained} to obtain: 
\begin{thm}[Equivalent to Theorem 4.3 in \cite{altman1999constrained}] \label{thm:LP}
 Let  $X_{\mathrm{ub}}$ be defined as in \eqref{eq:boundXmax}. The optimal stationary policy  $\{\xi_{x, r}^*\}_{x,r}$ 
  solving Problem 3 is given by
  \begin{equation}\label{eq:stat_optimal}
    \xi_{x,r}^*= \frac{\nu_{x, r}^*}{\mu_{x, r}^*},
    \end{equation}
 where $\nu_{x, r}^*$ and $\mu_{x, r}^*$ are obtained by the following LP problem:
   	\begin{subequations}
		\begin{align}
		\{\mu_{x, r}^*,&\nu_{x, r}^*\}=\arg\min_{\{\mu_{x,r},\nu_{x, r}\}}\sum_{x=1}^{X_{\mathrm{ub}}}\sum_{r=1}^R(\omega(r)x\mu_{x,r}\!+\!W\nu_{x, r}),\label{stationaryLPobj}\\
		\text{s.t.  }	&\mu_{1,r}=\sum_{x=1}^{X_{\mathrm{ub}}}\sum_{r'=1}^R\nu_{x, r'}P_{r',r},\label{stationaryLP1}\\
		&\mu_{x, r}= \sum_{r'=1}^R(\mu_{x-1, r'}-\nu_{x-1,r'})P_{r',r}, \forall x> 1,\label{stationaryLP2}\\
		&\sum_{x=1}^{X_{\mathrm{ub}}}\sum_{r=1}^R\mu_{x,r}=1, \label{stationaryLP3}\\
		&\nu_{x, r}\leq \mu_{x,r}, \label{stationaryLP4}\\
		&0\leq\mu_{x,r}      , 0\leq \nu_{x, r}, \forall x, r.\label{stationaryLP6} 
		\end{align}
		\label{LP}
	\end{subequations}       
\end{thm}


Before going further, we have some remarks:
\begin{itemize}
\item As mentioned in \cite[Theorem 4.3]{altman1999constrained}, any solution of the LP described in Theorem \ref{thm:LP} leads to a stationary optimal  policy through Eq.~\eqref{eq:stat_optimal}. Conversely, any stationary optimal policy for Problem 3 is also a solution to Theorem \ref{thm:LP}.
\item The set of constraints in the above-mentioned LP is just a straightforward application of  \cite[Theorem 4.3]{altman1999constrained} except  for Eq.~\eqref{stationaryLP4}. This constraint has been added since the LP is written with respect to the occupation measure and the steady-state distribution (which is a sum of all the occupation measures). So, by construction, $\nu_{x, r}\leq \mu_{x, r}$.
\end{itemize}

\subsection{An algorithmic solution for Problem 2}\label{subsec:problem2}

In Section \ref{subsec:problem3}, we described an LP approach to obtain an optimal policy $\pi_F^{n,*}(W)=\{\xi_{x,r}^{n,*}\}$ (or equivalently, $\{\mu_{x, r}^{n,*}(W), \nu_{x, r}^{n,*}(W)\}_{x,r}$) for Problem 3 for any file $n$ and Lagrange multiplier $W$. By Eq.~\eqref{eq:factorizedpol0}, the product of these policies minimizes the Lagrangian function  in Eq.~\eqref{eq:Lagrange} for the given multiplier $W$. So, at each time $t$,  for given AoI vector $(X_{1,t},\ldots, X_{N,t})= [x_1,\cdots, x_N]$  and popularity vector $(R_{1,t},\ldots, R_{N,t})=[r_1,\cdots, r_N]$, the optimal policy ${\boldsymbol \pi}_F^{*}(W)$ that minimizes the Lagrangian \eqref{eq:Lagrange} for parameter $W$ updates  each file $n$ independently of all the other files with probability $\xi^{n,*}_{x_n,r_n}(W)=\frac{\nu_{x_n, r_n}^{n,*}(W)}{\mu_{x_n, r_n}^{*,n}(W)}$.
The average proportion of time spent on downloading files in this optimal policy is
\begin{equation}\label{eq:constraint}
d^*(W):=\sum_{n=1}^N \sum_{x=1}^{X_{\mathrm{ub}}^n(W)}\sum_{r=1}^R\nu_{x, r}^{n,* } (W),
\end{equation}
and the expected average AoI 
\begin{equation}\label{eq:averageAoI}
 a^*(W)\triangleq \sum_{n=1}^N   \sum_{x=1}^{X_{\mathrm{ub}}^n(W)}\sum_{r=1}^R\omega(r)x\mu^{n,*}_{x, r}(W).
  	\end{equation}
Thus, for a given Lagrange multiplier $W$ and the optimal stationary policy $\boldsymbol \pi_F^{*}(W)$, the Lagrangian in \eqref{eq:Lagrange} can be compactly  written as:
\begin{equation}
\mathcal{L}(\boldsymbol \pi_F^{*}(W), W) =a^*(W)+ W d^*(W)-MW.
\end{equation}
It remains to find optimal value of $W$. As in \cite{mixing}, notice that for each $W$ and any policy $\pi$:
	\begin{equation}\label{eq:i}
	 a^*(W)+ W d^*(W)-MW \leq  a_{\boldsymbol\pi}+W d_{\boldsymbol\pi}-MW,
	\end{equation}
	where $a_{\boldsymbol\pi}$ and $d_{\boldsymbol\pi}$ denote the average proportion of time spent on downloading files and the expected average AoI under policy ${\boldsymbol\pi}$. By \eqref{eq:i}, if for some $\tilde{W}$ the constraint $d^*(\tilde{W})=M$, then it has the smallest expected AoI $a^*(\tilde{W})$ among all policies respecting constraint $d_{\boldsymbol\pi}\leq M$.   Since   $d^*(W)$ is  non-increasing in $W$ \cite[Lemma 3.3]{mixing},  if the desired value of $\tilde{W}$ above exists, it coincides with   \begin{equation}
	W^* \triangleq \inf\{W|d^*(W)\leq M\}. 
	\end{equation}
	In this case, the optimal policy for Problem~2 is given by $\boldsymbol{\pi}_R^{*}=\boldsymbol{\pi}_F^{*}(W^*)$.
	
	In case $W\mapsto d^*(W)$ is discontinuous around $W^*$, then we need to define two intermediate policies as follows. We recall that policy ${\boldsymbol\pi}_F^*(W)$ is a finite product of policies $\pi_F^{n,*}(W)$, each defined by the finite set $\{\xi_{x,r}^{n,*}(W)\}_{1\leq x\leq X_{\mathrm{ub}}^n(W), r\in\mathcal{R}}$. When we consider $W\in[0,W^*]$, according to (\ref{eq:boundXmax}), we can define   $X_{\mathrm{ub},\mathrm{ls}}^n= \max_{W\in[0,W^*]} X_{\mathrm{ub}}^n(W)$. We now consider the zero-padded set $\{\xi_{x,r}^{n,*}(W)\}_{1\leq x\leq X_{\mathrm{ub},\mathrm{ls}}, r\in\mathcal{R}}$. As each $\xi_{x,r}^{n,*}(W)$ is bounded by $1$, there exists a subsequence $\{W_{\mathrm{ls},\ell}\}_{\ell\geq 1}$ left-converging to $W^*$ s.t. $\xi_{x,r,\mathrm{ls}}^{n,*}:=\lim_{\ell\to \infty}  \xi_{x,r}^{n,*}(W_{\mathrm{ls},\ell})$ exists. Similarly, by considering $\overline{W}$ larger than $W^*$ and $X_{\mathrm{ub},\mathrm{rs}}^n= \max_{W\in[W^*,\overline{W}]} X_{\mathrm{ub}}^n(W)$, there exists a subsequence $\{W_{\mathrm{rs},\ell}\}_{\ell\geq 1}$ right-converging to $W^*$ s.t. $\xi_{x,r,\mathrm{rs}}^{n,*}:=\lim_{\ell\to\infty}\xi_{x,r}^{n,*}(W_{\mathrm{rs},\ell})$ exists. Consequently, with an abuse of notation (we do not mention the selected subsequences anymore as well as the integer index $\ell$), we can define
          \begin{equation}
	{\boldsymbol\pi}_{F,\mathrm{ls}}^* := \lim_{W \uparrow {W^*}}{\boldsymbol\pi}_F^*(W)
	\end{equation}
	and 
	\begin{equation} {\boldsymbol\pi}_{F,\mathrm{rs}}^* :=\lim_{W\downarrow {W^*}}{\boldsymbol\pi}_F^*(W).
        \end{equation}
        Notice that  ${\boldsymbol\pi}_{F,\mathrm{ls}}^*$ and ${\boldsymbol\pi}_{F,\mathrm{rs}}^*$  may depend on the selected subsequence but this has no impact on the final result of this subsection.

              As the mapping  ${\boldsymbol \pi} \mapsto d_{\boldsymbol\pi}$ is continuous,
              the following limits exist:
 \begin{equation}\label{eq:d_minus}
  d_{\mathrm{ls}}= \lim_{W\uparrow  {W^*}} d_{{\boldsymbol\pi}_{F}^*(W)}
  \end{equation}
and
\begin{equation}\label{eq:d_plus}
  d_{\mathrm{rs}}= \lim_{W\downarrow {W^*}}  d_{{\boldsymbol\pi}_{F}^*(W)}.
  \end{equation}

According to \eqref{eq:i}, see  \cite[Theorem 4.4]{mixing} and \cite{sennott93} for details,   the following mixed policy is optimal
  \begin{equation}\label{eq:piRstar}
	{\boldsymbol\pi}_R^*=\lambda {\boldsymbol\pi}_{F,\mathrm{ls}}^*+(1-\lambda){\boldsymbol\pi}_{F,\mathrm{rs}}^*,
\end{equation}
with
\begin{equation}
	\lambda :=\frac{M-d_{\mathrm{rs}}}{d_{\mathrm{ls}}-d_{\mathrm{rs}}}.
\end{equation}


Notice that the mixed policy in Eq.~\eqref{eq:piRstar} means that at time  $t=0$,  each of the two pure  policies  ${\boldsymbol\pi}_{F,\mathrm{ls}}^*$ and  ${\boldsymbol\pi}_{F,\mathrm{rs}}^*$ is chosen with probability $\lambda$ and  $1-\lambda$. The selected policy is then played until the end of the process in slot $t=T$. Such a policy is obviously not stationary, and moreover,  Constraint~\eqref{eq:bandwidthrelaxed} is not necessarily satisfied for all realizations of the decision process. An optimal \emph{stationary}  policy for 
Problem~2 can however easily be found based on ${\boldsymbol\pi}_R^*$ given above. The idea is to define the new steady-state probabilites  and occupation measures 
\begin{eqnarray}
\overline{\mu}_{x, r}^{n, *}& \triangleq&\lambda \mu_{x, r,\mathrm{ls}}^{n,*}+(1-\lambda)\mu_{x, r,\mathrm{rs}}^{n,*},\label{eq:barmu}\\
\overline{\nu}_{x, r}^{n, *}&\triangleq &\lambda \nu_{x, r,\mathrm{ls}}^{n,*}+(1-\lambda)\nu_{x, r,\mathrm{rs}}^{n,*},
\end{eqnarray}
where $\{\mu_{x, r,\mathrm{ls}}^{n,*}, \nu_{x, r,\mathrm{ls}}^{n,*}\}$ and  $\{\mu_{x, r,\mathrm{rs}}^{n,*}, \nu_{x, r,\mathrm{rs}}^{n,*}\}$ denote the steady-state probabilities and the occupation measures  associated with the policies  $\pi_{F,\mathrm{ls}}^{n,*}$ and $\pi_{F,\mathrm{rs}}^{n,*}$, respectively.
Consider now the corresponding stationary policy $\overline{\boldsymbol{\pi}}_{R}^*$,which updates at each time $t$ the file $n$ independently of all other files with a probability $\bar{\xi}_{x, r}^{n,*}\triangleq \frac{ \overline{\nu}_{x, r}^{n, *}}{\overline{\mu}_{x, r}^{n, *} }$, if $X_{n,t}=x$ and $R_{n,t}=r$.  By setting $\overline{X}_{\mathrm{ub}}^n = \max(X_{\mathrm{ub},\mathrm{ls}}^{n},X_{\mathrm{ub},\mathrm{rs}}^{n})$, one can show this policy achieves  the  following average AoI 
\begin{equation}\label{eq:age_final}
\sum_{n=1}^N\sum_{x=1}^{\overline{X}_{\mathrm{ub}}^{n}}  \sum_{r=1}^R   \omega(r)x[\lambda \mu_{x, r,\mathrm{ls}}^{n,*}  + (1-\lambda)\mu_{x, r,\mathrm{rs}}^{n,*}]
\end{equation}
and the following average downloading time
\begin{equation}\label{eq:constraint_final}
\sum_{n=1}^N\sum_{x=1}^{\overline{X}_{\mathrm{ub}}^{n}}  \sum_{r=1}^R [\lambda \nu_{x, r,\mathrm{ls}}^{n,*}  + (1-\lambda)\nu_{x, r,\mathrm{rs}}^{n,*}] =M.
\end{equation}
Consequently, this policy offers the same AoI  as the mixed policy $\boldsymbol{\pi}_{R}^*$ and satisfies the constraint, and is thus optimal.


%
%


\subsection{An algorithmic solution for Problem 1}\label{sec:algo}

 For our original Problem~1, we propose a (sub-optimal) policy $\hat{\boldsymbol\pi}$  that behaves as the  policy $\overline{\boldsymbol{\pi}}_{R}^*$ derived above,  except that at time-instances $t$ where $\overline{\boldsymbol{\pi}}_{R}^*$ downloads more than $M$ files, $\hat{\boldsymbol\pi}$  randomly choose $M$ files among those that were to be updated by strategy $\overline{\boldsymbol{\pi}}_{R}^*$.

\section{Optimality in an asymptotic regime }\label{sec:asymp}

As shown in the following theorem, for $N/M$  fixed and $N\to \infty$, the AoI of the practical policy
$\hat{\boldsymbol\pi}$ defined in Section \ref{sec:algo} converges   to the AoI of the optimal policy $\boldsymbol{\pi}^*$ that solves the original Problem 1.

\begin{thm}\label{thm:asymp}
  Assume that the $N$ files have the same popularity statistics $P_{r,r'}$ and the same expected number of requests given a popularity mode $\omega(r)$. Assume also a fixed ratio $\theta:=N/M$ independent of $N$. Assume that the Markov chain induced by policy $\overline{\boldsymbol\pi}_R^*$ is ergodic. The policy $\hat{\boldsymbol\pi}$ defined in Section \ref{sec:algo} is then asymptotically optimal in the sense
	\begin{equation}
		\lim_{N\rightarrow\infty\atop \theta=N/M}\frac{a_{\hat{\boldsymbol\pi}}-a_{{\boldsymbol\pi}^*}}{a_{{\boldsymbol\pi}^*}}=0,
	\end{equation}
where  ${\boldsymbol\pi}^*$ is the optimal policy for Problem 1.        

\end{thm}
\begin{IEEEproof}
Omitted for space limitation. See \cite{itw2021_comp}. 
	\end{IEEEproof}

Notice that the condition of ergodicity for the Markov chain is mild here since the randomness of the popularity mode may lead to aperiodic and positive recurrent states for the Markov chain induced by  $\overline{\boldsymbol\pi}_R^*$. 

Theorem \ref{thm:asymp} can  easily be extended to the case of a finite number of classes of users, where a class is formed by all the users with  the same  popularity statistics and the same expected number of requests given a popularity mode $\omega(r)$. The setup in Theorem~\ref{thm:asymp} corresponds to a single class.


\section{Numerical results}\label{sec:simus}
We consider two popularity modes $\mathcal{R}=\{1,2\}$ such that for some $q\in(0,1)$ all files $n$ have following transition matrix
\begin{equation}\label{eq:P}
  \mathbf{P}^n=\left[\begin{matrix}
q&1-q\\1-q&q
    \end{matrix}\right], \quad \forall n\in\{1,\ldots,N\}.
  \end{equation}

The expected number of requests of  file $n$ in the two states is $\omega_n(1)=0.2\overline{\omega}_n$ and $\omega_n(2)=1.8\overline{\omega}_n$, where  $\overline{\omega}_n\propto 1/n^\alpha$ follows a Zipf distribution with coefficient $\alpha=1.5$. Due to Eq.~(\ref{eq:P}), the steady-state probablity for both popularities are identical, so $\overline{\omega}_n$ is the average number of requests for file $n$.

\begin{figure}[htb]
	\centering
	\includegraphics[width=\columnwidth]{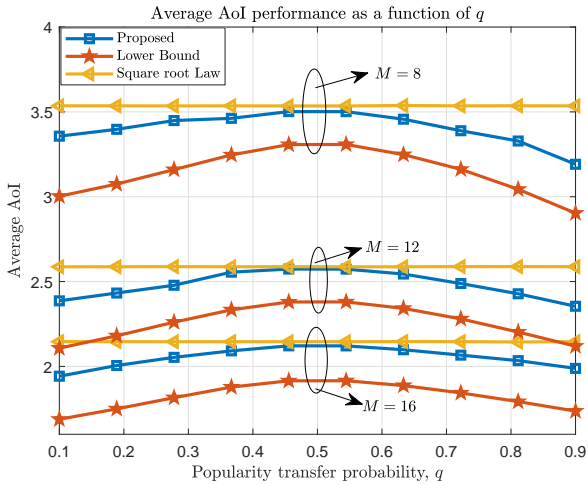}
	\caption{AoI vs $q$ for different $M$ and $N=64$.}
	\label{fig:vsQ}
\end{figure}

In Fig.~\ref{fig:vsQ}, we plot the average AoI versus $q$ with different $M$ and $N=64$ for the proposed policy $\hat{\boldsymbol{\pi}}$, the relaxed policy  $\overline{\boldsymbol{\pi}}_R^*$, and the square-root law \cite{yates17isit} designed with the average popularity. Notice that the AoI obtained with the relaxed policy $\overline{\boldsymbol{\pi}}_R^*$ is a lower bound. We also recall that the square-root law does not take into account the time-varying characteristic of the popularity mode.

 We first observe that the proposed policy achieves a better  AoI than the square-root law for any $q\neq 0.5$. For $q=0.5$  the current popularity mode does not provide any information on the next one, therefore the proposed policy provides marginal gain. When $q$ deviates more from $0.5$, the gain becomes larger since the future popularity mode is better predicted and it is thus more important to  take it into account.
Moreover, the square-root law does not depend on $q$ since the average number of requests over infinite horizon does not depend on $q$ in our simulations.
We further observe that when $M$ increases, the AoI decreases  since more updates can be done.
Finally, the gap between the proposed policy and the relaxed one (which does not satisfy the hard constraint (\ref{eq:C1})) slightly  depend on $q$.
\section{Conclusion}\label{sec:ccl}
The paper formulates a cache updating problem that aims  to minimize the average AoI of the requested files  under a bandwidth constraint on the server-cache link and assuming time-varying file popularities. We relaxed the hard bandwidth constraint and formulated the problem as a Constrained Markov Decision Process, which we then decoupled and solved through Linear Programming. Inspired by this approach, we proposed a practical updating strategy that satisfies the hard bandwidth constraint, and we showed that the proposed strategy is asymptotically optimal for a large number of files and under certain assumptions. Numerical results showed that for a fixed number of files the proposed strategy outperforms in many configurations the square-root law policy previously proposed for fixed popularities.

\bibliographystyle{ieeetr}
\bibliography{bibfile}

\end{document}


\title{Proof of Theorem 3 \\ in the paper entitled ``Minimizing the Age of Information of Time-Varying Requested File''} 

\author{
  \IEEEauthorblockN{Haoyue~Tang$^{1}$, Philippe Ciblat$^{2}$,  Jintao~Wang$^{1}$, Mich\`ele Wigger$^{2}$,~Roy Yates$^{3}$}
\IEEEauthorblockA{
	\textsuperscript{1}Beijing National Research Center for Information Science and Technology (BNRist),\\
	Dept. of Electronic Engineering, Tsinghua University, Beijing, China\\
	\textsuperscript{2}Telecom Paris, Institut Polytechnique de Paris, Palaiseau, France\\
	\textsuperscript{3}Rutgers University, New Brunswick, NJ, USA\\
        thy17@mails.tsinghua.edu.cn, \{ciblat, wigger\}@telecom-paris.fr, ryates@rutgers.edu}
}

\maketitle

\vspace{1cm}

We remind the Theorem we would like to prove.

\begin{thm}\label{thm:asymp}
  Assume that the $N$ files have the same popularity statistics $P_{r,r'}$ and the same expected number of requests given a popularity mode $\omega(r)$. Assume also a fixed ratio $\theta:=N/M$ independent of $N$. Assume that the Markov chain induced by policy $\overline{\boldsymbol\pi}_R^*$ is ergodic. The policy $\hat{\boldsymbol\pi}$ defined in Section II.D is then asymptotically optimal, i.e.,
	\begin{equation}
		\lim_{N\rightarrow\infty\atop \theta=N/M}\frac{a_{\hat{\boldsymbol\pi}}-a_{{\boldsymbol\pi}^*}}{a_{{\boldsymbol\pi}^*}}=0. 
	\end{equation}
where  ${\boldsymbol\pi}^*$ is the optimal policy for Problem 1.        

\end{thm}

\begin{IEEEproof}
	
	Recall that $a_{{\boldsymbol\pi}_R^*}$ or equivalently,  $a_{\overline{\boldsymbol{\pi}}_R^*}$ is the expected average AoI for Problem 2 with soft bandwidth constraint. As a consequence, we have  $a_{\overline{\boldsymbol\pi}_R^*} \leq a_{{\boldsymbol\pi}^*}$. Moreover  $a_{{\boldsymbol\pi}^*}<a_{\hat{\boldsymbol\pi}} $, it is then easy to check that 
	\begin{equation}
	0\leq \frac{a_{\hat{\boldsymbol\pi}}- a_{{\boldsymbol\pi}^*}}{  a_{{\boldsymbol\pi}^*} }\leq\frac{a_{\hat{\boldsymbol\pi}}-a_{\overline{\boldsymbol\pi}_R^*} }{ a_{\overline{\boldsymbol\pi}_R^*}}.\label{eq:relaxedUB}
	\end{equation}

        Given $W$, the LP associated with Theorem 2 is the same for each file since $\omega$ and $P_{r,r'}$ are now independent of $n$. So given $W$, $\{\mu_{x, r}^{n, *}(W), \nu_{x,r}^{n, *}(W)\}$ or equivalently $\{\xi_{x,r}^{n, *}(W)\}$ do not depend on $n$ and we omit the superscript $n$ since it plays no role. Then the Lagrangian of Eq.~(5) for the optimal policy can be written as
$$	\mathcal{L}({\boldsymbol\pi}_F^*(W), W)=N\left(  \sum_{x,r}^{X_{\mathrm{ub}}} \omega(r) x\mu_{x, r}^{ *}(W) +W \nu_{x,r}^{*}(W)  -W\frac{1}{\theta}\right). 
        $$
    
        Since $\theta$ is a constant independent of $N$, the optimization of  $W\mapsto \mathcal{L}({\boldsymbol\pi}_F^*(W), W)$ is independent of $N$. Consequently $\boldsymbol{\pi}_R^*$ or equivalently  $\overline{\boldsymbol{\pi}}_R^*$ is independent of $N$. Thus we also omit the optimal value of $W$ since it does not depend on $N$. As a result,  $\{\overline{\mu}_{x,r}^*, \overline{\nu}_{x,r}^*\}$ and $\{\overline{\xi}_{x,r}^*\}$ of policy $\overline{\boldsymbol{\pi}}_R^*$ are the same throughout the proof.

	Denote $\tau_r=\min_{x}\{\overline{\xi}_{x,r}^*>0\}$ as the minimum AoI that file $n$ may be updated at popularity state $r$ following policy $\overline{{\boldsymbol\pi}}_R^*$. Recall that $\overline{X}_{\mathrm{ub}}$ is an upper-bound for the largest updating threshold of $\overline{\boldsymbol\pi}_R^*$. Consequently, $\overline{\xi}_{\overline{X}_{\mathrm{ub}}, r}^*=1, \forall r$ and the AoI following policy $\overline{\boldsymbol\pi}_R^*$ cannot exceed $\overline{X}_{\mathrm{ub}}$ even at time $t=1$ since we initialize the age to be $1$. Denote $\Gamma=\overline{X}_{\mathrm{ub}}-\min_{r}\{\tau_r\}$ be the difference between the largest updating threshold and the smallest updating threshold of $\overline{\boldsymbol\pi}_R^*$. These thresholds are independent of $N$.  
	
Let $\mathcal{F}_t$ be the set of users the policy $\overline{\boldsymbol\pi}_R^*$ suggests to update at time $t$. According to policy $\overline{{\boldsymbol\pi}}_R^*$, each file $n$ belongs to $\mathcal{F}_t$ with probability $\overline{\xi}_{X_{n,t}, R_{n, t}}^*$. Then 
consider file $n$ in $\mathcal{ F}_t$ but is not updated by the policy $\hat{\boldsymbol\pi}$ in slot $t$. Since file $n\in\mathcal{F}_t$ implies $\overline{\xi}_{X_{n, t}, R_{n, t}}^*>0$ (if not, the file $n$ can not be put in $\mathcal{F}_t$ because it has been decided deterministically not to update it), the AoI in slot $t$ satisfies $X_{n,t}\geq\min_{r}\{\tau_r\}$. We then upper bound the probability that file $n$ is still not updated by policy $\hat{\pi}$ over the next $t'$ consecutive slots:
\begin{itemize}
\item If $t'\leq \Gamma$, with probability at most 1, the file is still not updated from slot $t$ to slot $t+t'$. 
\item If $t'>\Gamma$, then in slot $t+\Gamma$, its AoI $X_{n,t+\Gamma}\geq\min_r\{\tau_{r}\}+\Gamma\geq \overline{X}_{\mathrm{ub}}$. Thus in slot $k\in(t+\Gamma, t+t']$, if policy $\overline{\boldsymbol\pi}_R^*$ is used then file $n$ will be necessary updated since too old. So $n\in \mathcal{F}_k$.
  Let $\mathcal{D}_k$ be the set of users updated by policy $\hat{\boldsymbol\pi}$ at time $k$. By construction $\mathcal{D}_k\subseteq \mathcal{F}_k$. 
      The probability $\text{Pr}(n\notin\mathcal{D}_k|n\in\mathcal{ F}_k)=\frac{|\mathcal{F}_k|-M}{|\mathcal{F}_k|}\leq\frac{N-M}{N}=1-\theta$. Hence, the probability that file $n$ is still not updated after $t'$ consecutive slots can be upper bounded by $(1-\theta)^{(t'-\Gamma)}, \forall t'>\Gamma$.
\end{itemize}

In conclusion, for file $n\in\mathcal{ F}_t$ but not updated in slot $t$, the probability that file $n$ is still not updated after the next $t'$ consecutive slots can be upper bounded by $(1-\theta)^{(t'-\Gamma)^+}$, where $(\cdot)^+=\max\{\cdot, 0\}$. For each slot $k\in(t, t+t']$, the AoI of file $n$ can be computed by $X_{n, k}=X_{n, t}+(k-t)$. 
Thus, if file $n\in\mathcal{ F}_t$ but not scheduled in slot $t$ by policy $\hat{\boldsymbol\pi}$, the expected additional AoI $\delta_a(X_{n,t})$ compared to the case of an update in slot $t$ can be upper bounded by:
\begin{eqnarray*}
	\delta_a(X_{n,t})&\leq&\sum_{t'=1}^\infty(1-\theta)^{(t'-\Gamma)^+}\max_r\{\omega(r)\}(X_{n,t}+t')\nonumber\\
	&\leq&\max_r\{\omega(r)\}\left[\left(\Gamma-1+\frac{1}{\theta}\right)X_{n,t}+\frac{\Gamma(\Gamma-1)}{2}+\frac{1}{\theta}\Gamma+\frac{1}{\theta}\right]. 
\end{eqnarray*}

Next, we will upper bound the difference between $a_{\hat{\boldsymbol\pi}}$ and $a_{\overline{\boldsymbol\pi}_R^*}$.  For doing that, we consider a novel strategy, denoted by $\hat{\boldsymbol\pi}_S$ that behaves as $\hat{\boldsymbol{\pi}}$ except at time instances $t$ satisfy $|\mathcal{F}_t|>M$, policy $\hat{\boldsymbol\pi}_S$ updates all the files in set $\mathcal{ F}_t$ and adds a penalty $\delta_a(X_{n,t})$ for each file $n$ satisfies $n\in\mathcal{ F}_t/\mathcal{ D}_t$. As $M$ files are chosen uniformly to fill the set $\mathcal{ D}_t$, then if $|\mathcal{F}_t|> M$, each file in $\mathcal{F}_t$ is not selected into the set $\mathcal{ D}_t$ with probability $\frac{|\mathcal{F}_t|-M}{|\mathcal{F}_t|}$. The expected additional AoI assigned in slot $t$ can be thus upper bounded by:
\begin{equation}\label{eq:agemax}
\mathbbm{1}_{|\mathcal{F}_t|>M}\sum_{n=1}^N\mathbbm{1}_{n\in\mathcal{F}_t}\delta_a(X_{n,t})\frac{|\mathcal{F}_t|-M}{|\mathcal{F}_t|}\leq\mathbbm{1}_{|\mathcal{F}_t|>M}\sum_{n=1}^N\delta_a(X_{n,t})\frac{|\mathcal{F}_t|-M}{M}.
\end{equation}

The expected AoI obtained by $\hat{\boldsymbol\pi}_S$ is thus larger than those of $\hat{\boldsymbol\pi}$ since the file not updated under $\hat{\boldsymbol\pi}$  is updated under $\hat{\boldsymbol\pi}$ before achieving the extra age  $\delta_a(X_{n,t})$. Consequently 
$$a_{\hat{\boldsymbol\pi}} \leq a_{\hat{\boldsymbol\pi}_S}.$$
and also
$$0 \leq a_{\hat{\boldsymbol\pi}}-a_{\overline{\boldsymbol\pi}_R^*} \leq a_{\hat{\boldsymbol\pi}_S}-a_{\overline{\boldsymbol\pi}_R^*}.$$
Let $\overline{\mathcal{F}_t}= \mathbb{E}_{\overline{\boldsymbol\pi}_R^*}[|\mathcal{F}_t|]$ be also the expected average number of files updated in slot $t$ for policy $\overline{\boldsymbol\pi}_R^*$. We have

\begin{eqnarray}
 a_{\hat{\boldsymbol\pi}}-a_{\overline{\boldsymbol\pi}_R^*} &\leq& a_{\hat{\boldsymbol\pi}_S}-a_{\overline{\boldsymbol\pi}_R^*}\nonumber\\
&\overset{(a)}{\leq}&\lim_{T\rightarrow\infty}\frac{1}{T}\mathbb{E}_{\overline{\boldsymbol\pi}_R^*}\left[\sum_{t=1}^T\left( \mathbbm{1}_{|\mathcal{F}_t|>M}\sum_{n=1}^N\delta_a(X_{n,t})\frac{|\mathcal{F}_t|-M}{M}\right)\right]\nonumber\\
&{=}&\lim_{T\rightarrow\infty}\frac{1}{T}\mathbb{E}_{\overline{\boldsymbol\pi}_R^*}\left[\sum_{t=1}^T\left(\sum_{n=1}^N\delta_a(X_{n,t})\frac{(|\mathcal{F}_t|-M)^+}{M}\right)\right]\nonumber\\
&\overset{(b)}{\leq}&\lim_{T\rightarrow\infty}\frac{1}{MT}\mathbb{E}_{\overline{\boldsymbol\pi}_R^*}\left[\sum_{t=1}^T\sum_{n=1}^N\delta_a(X_{n,t})\left(\left(|\mathcal{F}_t|-\overline{\mathcal{F}_t}\right)^++\left(\overline{\mathcal{F}_t}-M\right)^+\right)\right]\nonumber\\
&\overset{(c)}{\leq}&\lim_{T\rightarrow\infty}\frac{1}{MT}\mathbb{E}_{\overline{\boldsymbol\pi}_R^*}\left[\sum_{t=1}^T\sum_{n=1}^N\delta_a(X_{n,t})\left(\left||\mathcal{F}_t|-\overline{\mathcal{F}_t}\right|+\left|\overline{\mathcal{F}_t}-M\right|\right)\right]\nonumber\\
&\overset{(d)}{\leq}&\lim_{T\rightarrow\infty}\frac{1}{MT}\mathbb{E}_{\overline{\boldsymbol\pi}_R^*}\left[\sum_{t=1}^T\sum_{n=1}^N\delta_a(\overline{X}_{\mathrm{ub}})\left(\left||\mathcal{F}_t|-\overline{\mathcal{F}_t}\right|+\left|\overline{\mathcal{F}_t}-M\right|\right)\right]\nonumber\\
&=&\frac{N}{M}\delta_a(\overline{X}_{\mathrm{ub}})\left(\lim_{T\rightarrow\infty}\mathbb{E}_{\overline{\boldsymbol\pi}_R^*}\left[\frac{1}{T}\sum_{t=1}^T\left||\mathcal{F}_t|-\overline{\mathcal{F}_t}\right|\right]+\lim_{T\rightarrow\infty}\mathbb{E}_{\overline{\boldsymbol\pi}_R^*}\left[\frac{1}{T}\sum_{t=1}^T\left|\overline{\mathcal{F}_t}-M\right|\right]\right),\label{eq:agebound}
\end{eqnarray}
where inequality $(a)$ holds because of (\ref{eq:agemax}),  inequality $(b)$ holds because $(x+y)^+\leq x^++y^+$, and $(c)$ holds because $(x)^+\leq|x|$. Inequality $(d)$ is obtained because the extra cost $\delta_a$ is an increasing function of the age and the AoI $X_{n,t}$ under $\overline{\boldsymbol\pi}_R^*$ can not exceed the largest threshold $\overline{X}_{\mathrm{ub}}$. 

We need now to analyze the behavior of both terms involved in (\ref{eq:agebound}). This is done through both following Lemmas. 
\begin{lemma}\label{lem:1}
  \begin{equation}
\lim_{T\rightarrow\infty}\mathbb{E}_{\overline{\boldsymbol\pi}_R^*}\left[\frac{1}{T}\sum_{t=1}^T\left|\overline{\mathcal{F}_t}-M\right|\right]=0. 
\end{equation}
\end{lemma}
and
\begin{lemma}\label{lem:2}
  \begin{equation}
\lim_{T\rightarrow\infty}\mathbb{E}_{\overline{\boldsymbol\pi}_R^*}\left[\frac{1}{T}\sum_{t=1}^T\left||\mathcal{F}_t|-\overline{\mathcal{F}_t}\right|\right]\leq\sqrt{N}. \label{asymptotic2}
\end{equation}
\end{lemma}
Proofs are given in Appendix.

Since $\omega(r)$ is fixed and $\overline{X}_{\mathrm{ub}}$ do not change with $N$, the extra cost $\delta_a(\overline{X}_{\mathrm{ub}})$ is finite and can be seen as a constant. As a result, we have
\begin{equation}
a_{\hat{\boldsymbol\pi}}-a_{\overline{\boldsymbol\pi}_R^*}\leq C_1\sqrt{N}.
\end{equation}
with
$$C_1=\frac{1}{\theta}\delta_a(X_{\mathrm{ub}})$$
a constant with respect to $N$.

In addition, we remind $a_{\overline{\boldsymbol\pi}_R^*} $ can be computed by:
\begin{equation}
a_{\overline{\boldsymbol\pi}_R^*}=\sum_{n=1}^N\sum_{r=1}^R\sum_{x=1}^{\overline{X}_{\mathrm{ub}}  }\omega(r)\overline{\mu}_{x, r}^{n, *}=N\left(\sum_{r=1}^R\sum_{x=1}^{\overline{X}_{\mathrm{ub}}}\omega(r)\overline{\mu}_{x, r}^{*}\right).
\end{equation}
The second equality holds since $\overline{\mu}_{x, r}^{n, *}$ does not depend on $n$ in our context.  In addition the term
$$C_2=\sum_{r=1}^R\sum_{x=1}^{\overline{X}_{\mathrm{ub}}}\omega(r)\overline{\mu}_{x, r}^{*}$$
is also a constant with respect to  $N$. 

Consequently, we get 
\begin{align}
 \frac{a_{\hat{\boldsymbol\pi}}-a_{\overline{\boldsymbol\pi}_R^*} }{ a_{\overline{\boldsymbol\pi}_R^*}} &\leq\frac{C_1 \sqrt{N}}{C_2N}=\mathcal{O}\left(\frac{1}{\sqrt{N}}\right). 
\end{align}

Finally considering \eqref{eq:relaxedUB}, we obtain
\begin{equation}
  \lim_{N\rightarrow\infty}\frac{a_{\hat{\boldsymbol\pi}}- a_{{\boldsymbol\pi}^*}}{  a_{{\boldsymbol\pi}^*} } =\lim_{N\rightarrow\infty}  \frac{a_{\hat{\boldsymbol\pi}}-a_{\overline{\boldsymbol\pi}_R^*} }{ a_{\overline{\boldsymbol\pi}_R^*}}=0.
\end{equation}
\end{IEEEproof}

\appendix
\subsection{Proof of Lemma \ref{lem:1}}
We consider the policy $\overline{\boldsymbol\pi}_R^*$ is applied.
As the Markov chain is considered as ergodic  with steady state distribution $\overline{\mu}_{x,r}^{*}$ for state $(x,r)$ whatever $n$ and $N$, we know that for every $\varepsilon$, there exists $T_\varepsilon$ such that for any $t>T_\varepsilon$, we have:
\begin{equation}
\left|\text{Pr}(X_{n,t}=x, R_{n,t}=r)-\overline{\mu}_{x,r}^{ *}\right|\leq \epsilon, \forall 1\leq x\leq \overline{X}_{\mathrm{ub}},
\label{eq:MCstable}
\end{equation}
and for $x>\overline{X}_{\mathrm{ub}}$ we have:
\begin{equation}\label{eq:P0}
\text{Pr}(X_{n,t}=x, R_{n,t}=r)=0.
\end{equation}

Let $\text{Pr}(u_{n,t}=1)$ be the probability that file $n$ is updated in slot $t$.

For $\forall t>T_\varepsilon$, we have
\begin{eqnarray}
\text{Pr}(u_{n,t}=1)&=&\sum_{r=1}^R\sum_{x=1}^{\infty}\text{Pr}(X_{n,t}=x,R_{n,t}=r)\text{Pr}(u_{n,t}=1|X_{n,t}=x, R_{n,t}=r)\nonumber\\
&\overset{(a)}{=}&\sum_{r=1}^R\sum_{x=1}^{\overline{X}_{\mathrm{ub}}}\text{Pr}(X_{n,t}=x, R_{n,t}=r)\overline{\xi}_{x,r}^{*}\nonumber\\
&=&\sum_{r=1}^R\sum_{x=1}^{\overline{X}_{\mathrm{ub}}}\overline{\mu}_{x,r}^{*}\overline{\xi}_{x,r}^{*}+\sum_{r=1}^R\sum_{x=1}^{\overline{X}_{\mathrm{ub}}}\left(\text{Pr}(X_{n,t}=x, R_{n,t}=r)-\overline{\mu}_{x, r}^{*}\right)\overline{\xi}_{x,r}^{*}.\label{eq:updatedeviation}
\end{eqnarray}
where equality $(a)$ holds because of (\ref{eq:P0}), and $\overline{\xi}_{x,r}^*$ is the probability to update under policy $\overline{\boldsymbol\pi}_R^*$.

Eq.~\eqref{eq:updatedeviation} implies that, for $t\geq T_\varepsilon$, 
\begin{eqnarray}
\left|\text{Pr}(u_{n,t}=1)-\left(\sum_{r=1}^R\sum_{x=1}^{\overline{X}_{\mathrm{ub}}}\overline{\mu}_{x,r}^{*}\overline{\xi}_{x,r}^{*}\right)\right|&=&\left|\sum_{r=1}^R\sum_{x=1}^{\overline{X}_{\mathrm{ub}}}\left(\text{Pr}(X_{n,t}=x, R_{n,t}=r)-\overline{\mu}_{x, r}^{*}\right)\overline{\xi}_{x,r}^{*}\right|\nonumber\\
&\overset{(a)}{\leq}&\varepsilon R\overline{X}_{\mathrm{ub}},\label{eq:up}
\end{eqnarray}
where inequality $(a)$ is obtained by \eqref{eq:MCstable}.
%

It is easy to check that
$$\overline{\mathcal{F}_t}= \sum_{n=1}^N \mathbb{E}_{\overline{\boldsymbol{\pi}}_R^*}[u_{n,t}]=  \sum_{n=1}^N \text{Pr}(u_{n,t}=1).$$
and we remind that
\begin{equation}
\sum_{n=1}^N\sum_{r=1}^R\sum_{x=1}^{\overline{X}_{\mathrm{ub}}}\overline{\mu}_{x,r}^{*}\overline{\xi}_{x,r}^{*}=M.
\end{equation}
Consequently, for $t> T_\varepsilon$,  we have 
\begin{eqnarray*}
  \left| \overline{\mathcal{F}_t}-M\right| &= &\left|\sum_{n=1}^N \left(\text{Pr}(u_{n,t}=1)- \sum_{r=1}^R\sum_{x=1}^{\overline{X}_{\mathrm{ub}}}\overline{\mu}_{x,r}^{*}\overline{\xi}_{x,r}^{*}\right)  \right| \\
  &\leq & \sum_{n=1}^N \left|\text{Pr}(u_{n,t}=1)- \sum_{r=1}^R\sum_{x=1}^{\overline{X}_{\mathrm{ub}}}\overline{\mu}_{x,r}^{*}\overline{\xi}_{x,r}^{*}  \right|\\
  &\overset{\text{Eq.}~(\ref{eq:up})}{\leq}& N \varepsilon R\overline{X}_{\mathrm{ub}}
\end{eqnarray*}


Finally, we obtain
\begin{eqnarray*}
\lim_{T\rightarrow\infty}\mathbb{E}_{\overline{\boldsymbol\pi}_R^*}\left[\frac{1}{T}\sum_{t=1}^T|\overline{\mathcal{ F}_t}-M|\right]&=&\lim_{T\rightarrow\infty}\mathbb{E}_{\overline{\boldsymbol\pi}_R^*}\left[\frac{1}{T}\left(\sum_{t=1}^{T_\varepsilon}|\overline{\mathcal{ F}_t}-M|+\sum_{t=T_\varepsilon+1}^{T}|\overline{\mathcal{ F}_t}-M|\right)\right]\nonumber\\
&\leq&\lim_{T\rightarrow\infty}\left(\frac{T_\varepsilon N}{T}+\frac{T-T_\varepsilon}{T}\varepsilon NR\overline{X}_{\mathrm{ub}}\right)\nonumber\\
&=&\varepsilon  NR\overline{X}_{\mathrm{ub}}. 
\end{eqnarray*}
Previous manipulations just corresponds to Cesaro's sum.

By taking $\varepsilon$ arbitrarily small, we conclude the proof.

\subsection{Proof of Lemma \ref{lem:2}}
We have 
\begin{eqnarray}
  \mathbb{E}_{\overline{\boldsymbol\pi}_R^*}\left[\left|\vert\mathcal{F}_t\vert-\overline{\mathcal{ F}_t}\right|\right]&\overset{(a)}{\leq}&\sqrt{\mathbb{E}_{\overline{\boldsymbol\pi}_R^*}\left[\left||\mathcal{F}_t|-\overline{\mathcal{ F}_t}\right|^2\right]}\nonumber\\
 &\overset{(b)}{=}&\sqrt{\text{var}\left[\left|\mathcal{F}_t\right|\right]}.\label{eq:avgbias}
\end{eqnarray}
Inequality $(a)$ holds because of Jensen's inequality. Equality $(b)$ holds by definition $\overline{\mathcal{F}_t}=  \mathbb{E}_{\overline{\boldsymbol\pi}_R^*}[|\mathcal{F}_t|]$. 


Remind that $|\mathcal{F}_t|=\sum_{n=1}^Nu_{n,t}$. As the policy $\overline{\boldsymbol\pi}_R^*$ leads to separate independent policy for each file (the update decision for file $n$ depends only on its current state $(X_{n,t}, R_{n,t})$. Thus, the update decisions $\{u_{n,t}\}_n$ are independent binary random variables. Consequently 
\begin{equation}\label{eq:FB}
\text{var}[|\mathcal{ F}_t|]=\sum_{n=1}^N\text{var}[u_{n,t}]\overset{(a)}{\leq} N.
\end{equation}
Inequality $(a)$ holds since $u_{n,t}\leq 1$ for any $n,t$. 

Plugging \eqref{eq:FB} into \eqref{eq:avgbias} implies that
\begin{equation*}
\mathbb{E}_{\overline{\boldsymbol\pi}_R^*}\left[\left||\mathcal{ F}_t|-\overline{\mathcal{ F}_t}\right|\right]\leq\sqrt{N}, 
\end{equation*}
which concludes the proof.